# Machine-learning-informed parameter estimation improves the reliability of spinal cord diffusion MRI

Ting Gong[1*], Francesco Grussu[2,3], Claudia A. M. Gandini Wheeler-Kingshott[3,4,5], Daniel C Alexander[1], Hui Zhang[1]

1 Centre for Medical Image Computing & Department of Computer Science, University College London, London, United Kingdom

2 Radiomics Group, Vall d'Hebron Institute of Oncology, Vall d'Hebron Barcelona Hospital Campus, Barcelona, Spain

3 NMR Research Unit, Queen Square MS Centre, Queen Square Institute of Neurology, Faculty of Brain Sciences, University College London, London, United Kingdom

4 Brain Connectivity Centre, IRCCS Mondino Foundation, Pavia, Italy

5 Department of Brain and Behavioural Sciences, University of Pavia, Pavia, Italy

* Corresponding author

Corresponding email address: ting.gong@ucl.ac.uk

## Abstract

**Purpose**: We address the challenge of inaccurate parameter estimation in diffusion MRI when the signal-to-noise ratio (SNR) is very low, as in the spinal cord. The accuracy of conventional maximum-likelihood estimation (MLE) depends highly on initialisation. Unfavourable choices could result in suboptimal parameter estimates. Current methods to address this issue, such as grid search (GS) can increase computation time substantially.

**Methods**: We propose a machine learning (ML) informed MLE approach that combines conventional MLE with ML approaches synergistically. ML-based methods have been developed recently to improve the speed and precision of parameter estimation. However, they can generate high systematic bias in estimated parameters when SNR is low. In the proposed ML-MLE approach, an artificial neural network model is trained to provide sensible initialisation for MLE efficiently, with the final solution determined by MLE, avoiding biases typically affecting pure ML estimations.

**Results**: Using parameter estimation of neurite orientation dispersion and density imaging as an example, simulation and *in vivo* experiments suggest that the ML-MLE method can reduce outlier estimates from conventional MLE in white matter voxels affected by CSF contamination. It also accelerates computation compared to GS-MLE.

**Conclusion**: The ML-MLE method can improve the reliability of parameter estimation with reduced computation time compared to GS-MLE, making it a practical tool for diffusion dataset with low SNR.

## 1 Introduction

Being a non-invasive tool to characterise the microstructure of neural tissue in the central nervous system, diffusion MRI has been widely used to study the brain and is increasingly used to examine the spinal cord. In the spinal cord, most diffusion MRI studies have focused on using apparent diffusion coefficient (1) or diffusion tensor imaging (2) to evaluate the axonal integrity following pathological changes, such as in spinal cord injury (3–5), multiple sclerosis (6–8) and amyotrophic lateral sclerosis (9–11). In recent years, more studies have explored and applied advanced diffusion methods to investigate neuronal morphology-related microstructural properties in the spinal cord (12,13). These methods provide novel biomarkers for characterising microstructural alterations in spinal cord pathology, such as Neurite orientation dispersion and density imaging (NODDI) (14) applied to multiple sclerosis(15–17). Nevertheless, compared to their applications in the brain, the exploration of such methods in the spinal cord is still limited due to the difficult imaging environment of the spine (18,19) and its limited size. Though advanced spinal cord MRI has experienced tremendous advances both in terms of image acquisition (20,21) and analysis (22,23), the challenge of intrinsic lower signal-to-noise ratio (SNR) than typically seen in the brain remains.

The low SNR in spinal cord DWIs can lead to inaccurate parameter estimation, challenging the application of diffusion methods - especially advanced ones - in spinal cord studies. When fitting a microstructure model to measurements of low SNR, a conventional method such as the maximum likelihood estimation (MLE) is often used with numerical optimisation (24). The MLE method finds the best estimate that gives the highest likelihood of the measurements under an appropriate noise model. This procedure of finding such a best estimate generally involves an iterative algorithm starting from some initial guess in the parameter space. Inaccurate estimation can happen when the optimisation is stopped at a suboptimal location. In this case, choosing a suitable starting point for nonlinear optimisation can be crucial to ensure convergence to the correct solution.

Several approaches have been developed to improve conventional fitting, which however are usually performed voxel by voxel and therefore could increase computation time significantly. One such method is to conduct a grid search (GS) in the parameter space before optimisation to find the initial guess that is more likely to give the best likelihood (25,26). Other more time-consuming approaches include the multi-start method that repeats the optimisation multiple times with different starting points and stochastic methods such as simulated annealing (27) and full Markov-Chain Monte Carlo (28). Besides non-linear optimisation methods, linearisation is used in some methods to mitigate the starting point problem at the cost of possible approximation errors (29).

Recently, machine learning (ML)-based techniques have been developed as an alternative to the conventional estimation approach in diffusion MRI (30–36). These methods are known for their speed and precision. A trained ML model can estimate microstructure parameters in large datasets almost instantly. However, a recent study suggests ML estimation can generate systematic bias in estimated parameters especially when SNR is low (37). This bias could hinder the applicability and interpretability of ML-based methods in clinical settings.

To address the challenge of parameter estimation under low SNR, we propose an ML-informed MLE (ML-MLE) approach that combines the conventional MLE and the ML approaches synergistically. The approach initialises the MLE efficiently and optimally by training an artificial neural network (ANN) model. Suitable initialisations can be identified instantly for large datasets through a network inference step, which saves computation time compared to GS-MLE. At the same time, the final solutions are determined by MLE, avoiding biases typically affecting pure ML estimation.

## 2 Theory

This section describes of the conventional MLE method, and the GS procedure to find the starting point for MLE optimisation.

### 2.1 MLE

The MLE method finds the parameter estimate that maximises the likelihood of the data we measure under a statistical model of the noise. This is mathematically described as:

$$\widetilde{\boldsymbol{\theta}} = arg\max_{\boldsymbol{\theta}} p(\boldsymbol{y}|\boldsymbol{\theta}) \tag{1}$$

where $\boldsymbol{y} = [y_1, \dots, y_N]$ is the vector of measured signals and $\boldsymbol{\theta}$ the vector of parameters of interest, and $p(\boldsymbol{y}|\boldsymbol{\theta})$ is the probability density of observing $\boldsymbol{y}$ given $\boldsymbol{\theta}$. Under the typical assumption that the noise on each measurement is independent and identical,

$$\widetilde{\boldsymbol{\theta}} = arg\max_{\boldsymbol{\theta}} \sum_{n=1}^{N} \log p(y_n|\boldsymbol{\theta}) \tag{2}$$

Given that the magnitude of MR signals is independently Rician distributed(38), the probability density of observing $y_n$ given $\boldsymbol{\theta}$ can then be expressed as(24):

$$p(y_n|\boldsymbol{\theta}) = \frac{y_n}{\sigma^2} e^{\left(-\frac{y_n^2 + S(b_n, \boldsymbol{g_n}; \boldsymbol{\theta})^2}{2\sigma^2}\right)} I_0\left(\frac{y_n S(b_n, \boldsymbol{g_n}; \boldsymbol{\theta})}{\sigma^2}\right) \tag{3}$$

where $S(b_n, \boldsymbol{g_n}; \boldsymbol{\theta})$ is the noise-free signal predicted by the forward model with the diffusion sensitising factor $b_n$, gradient direction $\boldsymbol{g_n}$ and tissue parameter $\boldsymbol{\theta}$; $\sigma$ is the standard deviation of noise level and could be approximated by the standard deviation of S(b=0) measurements. $I_0$ is the modified Bessel function of the first kind with order zero.

To solve this problem, choosing a good starting guess for $\boldsymbol{\theta}$ can be crucial. $\boldsymbol{\theta}$ is then optimised iteratively using non-linear optimisation until the maximum log-likelihood is found under some stopping criterion of choice.

### 2.2 Grid search for choosing the starting point

In a grid search method, the best starting point is chosen by firstly computing the log-likelihood of a set of locations in the parameter space, and then comparing the likelihoods and setting the starting location as the parameter combination that gives the maximum likelihood. The set of locations is chosen to reside on a regular grid in the parameter space. Each parameter is allowed to take a set of evenly spaced values within its plausible range (26).

The GS process of finding the starting point can be very time-consuming, and the time increases as the number of voxel increases. Depending on the dimension of the parameter space, i.e., the number of parameters to estimate in the model ($n$), and the number of values sampled for each parameter for searching ($N_k$), the number of evaluations required for each voxel is $N = \prod_{k=1}^{n} N_k$, which grows quickly as the dimension of parameter space increases.

## 3 Methods

This section describes the proposed ML-informed method for finding the starting point for MLE, followed by implementation details and experiments for assessment of the ML-MLE method.

### 3.1 ML-informed method

With ML-informed method, the starting locations of a large dataset can be generated directly and efficiently with an ANN model. The ANN model is designed to map the diffusion signals to the diffusion model-derived parameters. The training of such a model is performed on a simulated dataset with known ground truth and then applied to the target dataset to get starting points close to optimal. The simulated training dataset is generated using a forward diffusion model with uniformly sampled tissue parameters (37) and the same imaging protocol as the target dataset. Once the model is trained on the simulated dataset, it can be applied instantly to any datasets with the same diffusion acquisition protocol to get the starting points.

### 3.2 Implementation details

The NODDI model (14) is investigated as a demonstration of estimating advanced diffusion parameters in the spinal cord. The parameters of interest include intra-neurite fraction $f_{in}$, orientation dispersion index $ODI$, and free water fraction $f_{iso}$. As the fibre bundle in the healthy spinal cord is highly aligned with the superior-inferior direction of the body, the fibre orientation in the model is set to the principal direction estimated from diffusion tensor (2) before estimating the 3 parameters of interest.

For the ML estimation of ML-MLE method, the ANN follows a standard architecture (33,39). It contains an input layer with the number of channels equal to the number of DWI volumes including the b=0 signal, 3 hidden layers with 150 units in each, and an output layer for the 3 NODDI parameters. The loss function is defined as the mean square error of the target parameters. The rectified linear unit is used as the activation function for all the hidden layers and Sigmoid is used for the output layer to guarantee the ranges of output parameters are between 0 and 1.

The training dataset is generated synthetically with the NODDI model from known tissue parameter values of $f_{in}, ODI$ and $f_{iso}$, all ranging from 0 to1, with $f_{in}$ and $ODI$ uniformly distributed to achieve lower estimation bias (37); the distribution for $f_{iso}$ contains more samples below 0.4 as including high $f_{iso}$ samples for training will bias other model parameters and $f_{iso}$ from in vivo dataset are generally below 0.3 in the cord GM and WM (supplementary materials M1). A total of one million samples are generated for training. The same diffusion sampling scheme as the target datasets is used to synthesise the signal $S$. Rician noise is added to the simulated signal by $S_{rician} = \sqrt{(S + N_1(0,\sigma)^2 + N_2(0,\sigma^2)}$, where $\sigma$ is the standard deviation of Gaussian noise level. The SNR levels for in vivo acquisition vary among different segments of the cord and are typically below 10 when estimated using b=0 signals (40). For the training datasets, the SNR can be chosen to be the value estimated from the target in vivo datasets. This is the strategy adopted here, resulting in an SNR level of 10 to demonstrate the method. Because b=0 signals naturally vary between voxels in in vivo datasets, the evaluation with in vivo data enables the generalisation of the trained model to be evaluated for different SNRs.

### 3.3 Experiments

The ML-MLE method is evaluated in terms of computation speed and accuracy, and compared to GS-MLE and direct ML estimation. Specifically, the computation time is compared between ML-MLE and GS-MLE with simulation and *in vivo* datasets. The accuracy and precision of estimation are compared for all the methods with simulation datasets. Finally, the findings from the simulation are further demonstrated on *in vivo* datasets.

For the comparison to the conventional method, GS-MLE is implemented based on the NODDI MATLAB toolbox on the *in vivo* and simulated datasets. The grids for searching are uniformly sampled from 0 to 1 with a separation of 0.25 for $f_{in}$, $ODI$, and $f_{iso}$, resulting in 125 locations in the parameter space. The location that gives the highest log-likelihood is used as the starting point for MLE.

#### 3.3.1 *In vivo* data

*In vivo* spinal cord data from a previous study (12) were retrospectively analysed. These consisted of scans acquired with a 3T Philips Achieva scanner from 5 healthy subjects. A multi-shell diffusion protocol optimised for NODDI parameter estimation was used for data acquisition (14): 30 and 60 diffusion gradient directions were applied respectively for the first shell at b=711 s/mm$^2$ and the second shell at b=2855 s/mm$^2$; 6 repetitions of b = 0 images were interleaved through the whole session. Scans were performed axial-oblique by carefully aligning with the slice-selection direction (z) on a sagittal localiser. All images underwent pre-processing steps for motion correction and tissue segmentation of white matter (WM) and grey matter (GM). Details about acquisition parameters and data pre-processing can be found in the original study (12). The SNR levels of the datasets are measured and reported in supplementary materials M2.

#### 3.3.2 Simulation

Diffusion measurements are simulated for quantitative comparison of all methods. Noise-free signals $S(b_n, \boldsymbol{g_n}; \boldsymbol{\theta})$ are synthesised using the NODDI model with the same diffusion imaging protocol $(b_n, \boldsymbol{g_n})$ as the *in vivo* data acquisition, and with typical sets of tissue parameters ($\boldsymbol{\theta}$) suggested in the previous study (12). Specifically, tissue parameters of $f_{in} = [0.35, 0.45, 0.55, 0.65, 0.75]$, $ODI = [0.02, 0.12, 0.22]$ are simulated. Different levels of CSF contamination are explored with $f_{iso} = [0, 0.1, 0.2, 0.3]$. For each combination of tissue parameters, noisy measurements are generated 100 times by adding random Rician noise to the corresponding noise-free signals as described in the training dataset. An SNR level of 10 for the b=0 signal is assessed.

## 4 Results

### 4.1 Computation time

The computation time for the ML-MLE (about 175 s/ 1000 voxels and 215 s/ 1000 voxels for simulated and *in vivo* dataset) is about 1.75 times faster than the GS-MLE (about 300 s/ 1000 voxels and 400 s/ 1000 voxels for simulated and *in vivo* dataset). The training time for the ANN model is about 7 mins on a single CPU. Once the model is trained, it is applied to all new datasets including simulated and *in vivo* data to get the initialisation for MLE. The application of the model to new datasets is completed almost instantly.

### 4.2 Simulation: accuracy and precision of estimation

Figure 1 demonstrates the joint distribution of $f_{in}$, and $f_{iso}$ from noisy simulations for each method in the WM (ODI =0.02). While the MLE-based method can give accurate estimates of $f_{in}$ and $f_{iso}$ in most cases, the $f_{in}$ estimation from GS-MLE was stuck at an outlier value of 1 for certain noise realisations in all combinations of tissue parameters (Figure 1.A). These outliers, however, are eliminated in the ML-informed MLE method (Figure 1.B), therefore improving the accuracy and precision of estimation especially when there is high CSF contamination. The estimates directly from the ML method improve the precision significantly but induce systematic biases in the parameters (Figure 1.C). Specifically, $f_{iso}$ is overestimated in simulations with low CSF contamination and underestimated in simulations with high CSF contamination; $f_{in}$, is underestimated in simulations with high CSF contamination. Distributions for a wider range of $f_{in}$ can be found in supplementary materials M3.

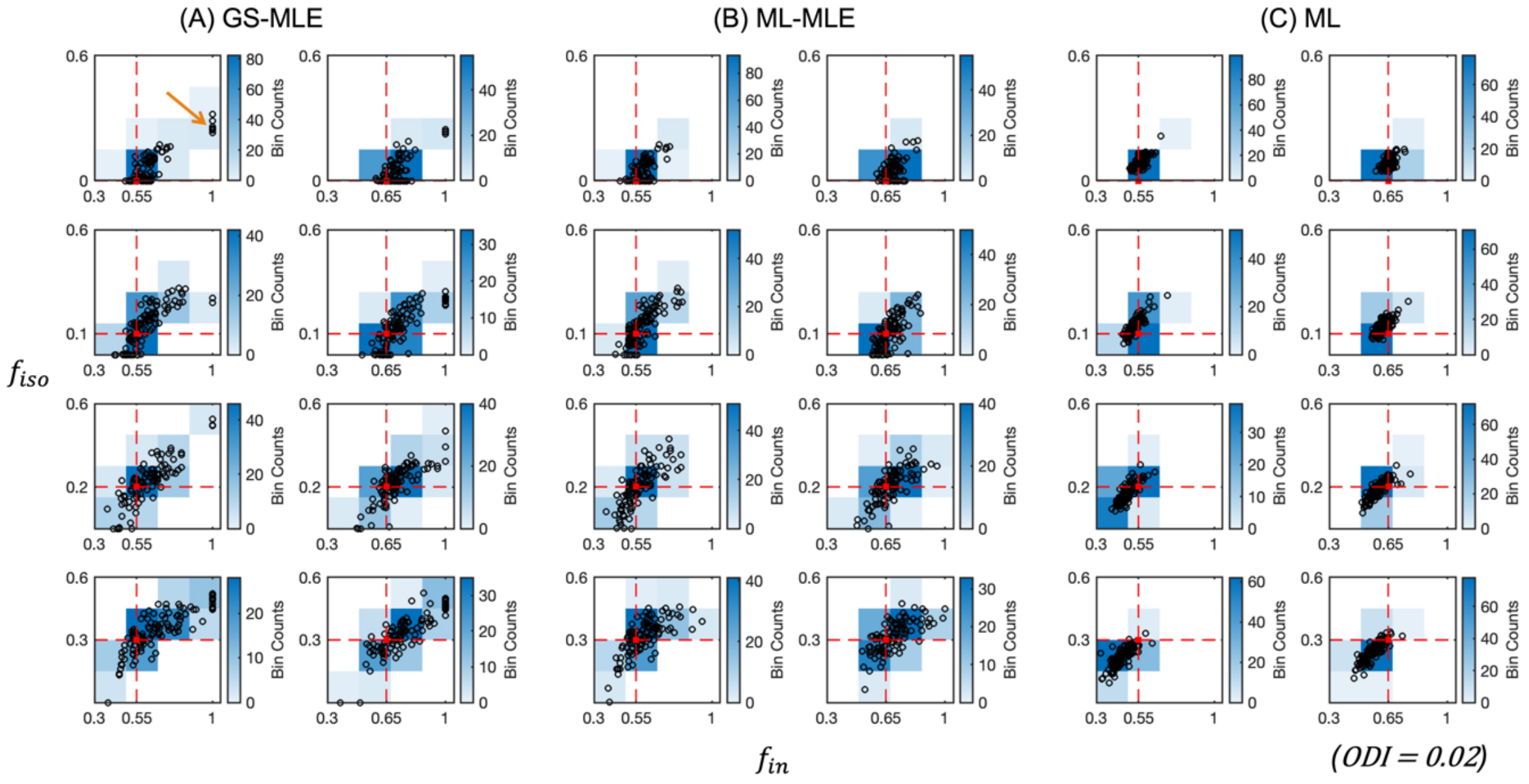

Figure 1. 2D distribution of estimated intra-neurite fraction $f_{in}$ and free water fraction $f_{iso}$ from noisy simulation of WM with (A) GS-MLE, (B) ML-MLE and (c) ML estimation. For each set of tissue parameters, the ground truth is marked as a red square. An example of outlier estimates of $f_{in}$ from GS-MLE is indicated by the orange arrow. ML estimation generates a negative bias in the estimated $f_{in}$ when there is CSF contamination, which is typical in the WM of the spinal cord.

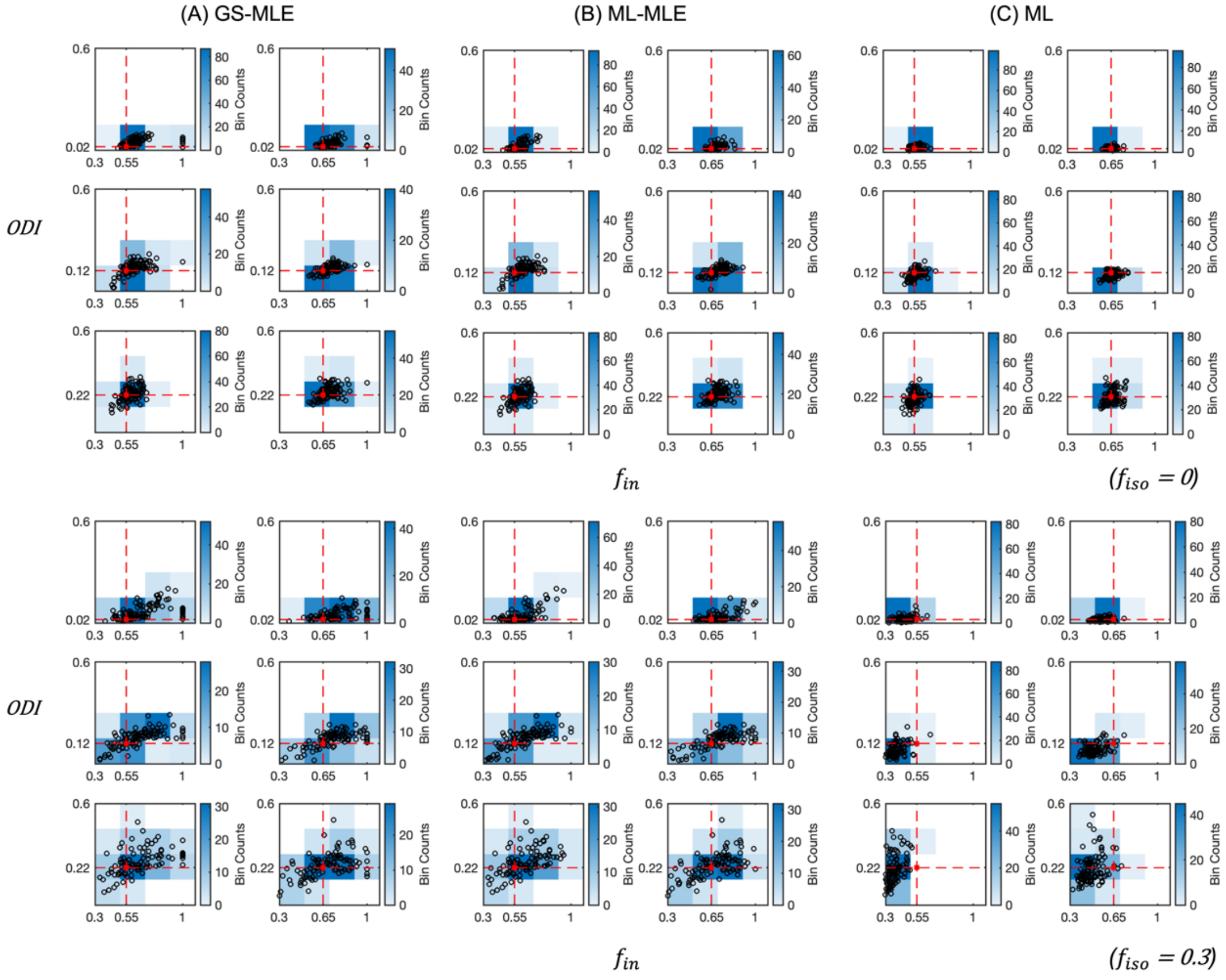

Figure 2. 2D distribution of estimated intra-neurite fraction $f_{in}$ and orientation dispersion index ODI from noisy simulation with (A) GS-MLE, (B) ML-MLE and (c) ML estimation. For each set of tissue parameters, the ground truth is marked as a red square. The estimation accuracy and precision of ODI is less affected by CSF contamination compared to $f_{in}$ for all methods.

Figure 2 demonstrates the joint distribution of $f_{in}$ and ODI from noisy simulations for each method without ($f_{iso} = 0$) and with high CSF contamination ($f_{iso} = 0.3$). Compared to $f_{in}$ and $f_{iso}$ estimation, ODI estimation is less affected by the noise or methods used. When ODI is low as in WM (ODI =0.02), its estimate precision is higher than those with high ODI, especially for direct ML estimation.

#### 4.3 Robust *in vivo* estimation

The *in vivo* results agree well with the simulation findings. Figure 3 shows parameter maps from example slices of a single subject. The ML-MLE gives an overall similar estimation to GS-MLE in the GM but eliminates most of the outlier estimates of $f_{in}$ in the WM, likely affected by CSF contamination indicated by high $f_{iso}$. The ML estimated $f_{in}$, are systematically lower than GS-MLE; the ML estimated $f_{iso}$ are systematically higher than GS-MLE in GM regions with lower CSF contamination.

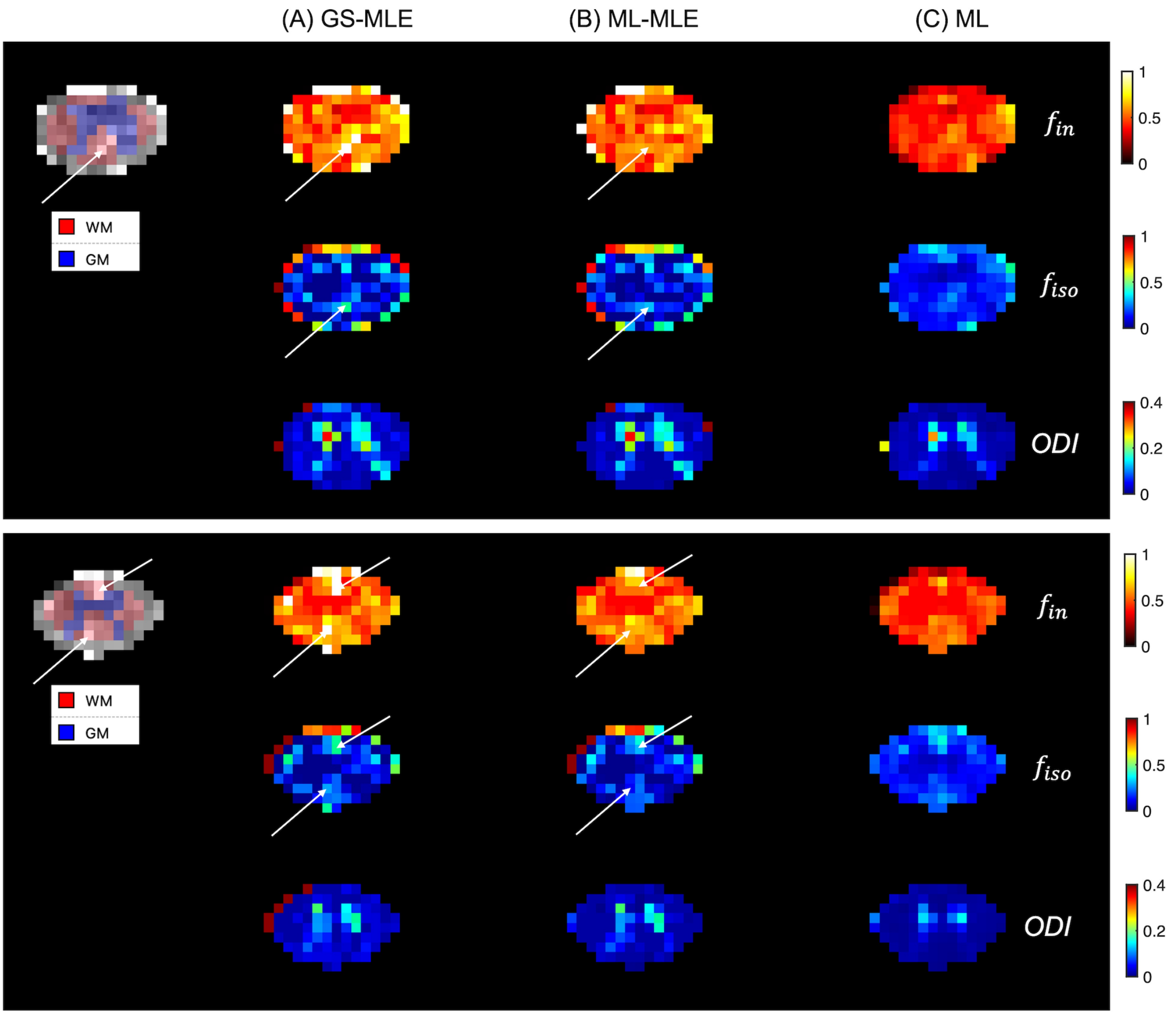

Figure 3. Example image slices of estimated $f_{in}$, $f_{iso}$ and ODI maps from a typical subject from different methods (A-C); WM and GM masks are overlayed on $f_{in}$ maps estimated with GS-MLE. Outliers in the cord from the GS-MLE method are mostly in the WM as indicated by the white arrow. ML-MLE reduces these outliers in WM while giving similar estimation in GM compared to GS-MLE.

These findings are consistent across all subjects with varying SNR levels. Figure 4 shows the distribution of $f_{in}$ estimation in both WM and GM. Outliers estimates of $f_{in}$, are found for GS-MLE estimation for all subjects, which are largely eliminated by ML-MLE, bringing closer the mean and median values of the distribution. ML estimation while improving the precision, gives systematically lower estimates of $f_{in}$, in all subjects. Table 1 gives the summary of mean and standard deviations of parameter estimates in WM and GM from all the subjects. With the elimination of outliers from ML-MLE in the WM, the standard deviations of $f_{in}$ estimates are lower than the GS-MLE method, indicating improved precision. In GM, the two methods give similar mean values and standard deviations. ML estimation gives systematically lower mean estimates of $f_{in}$, in all subjects.

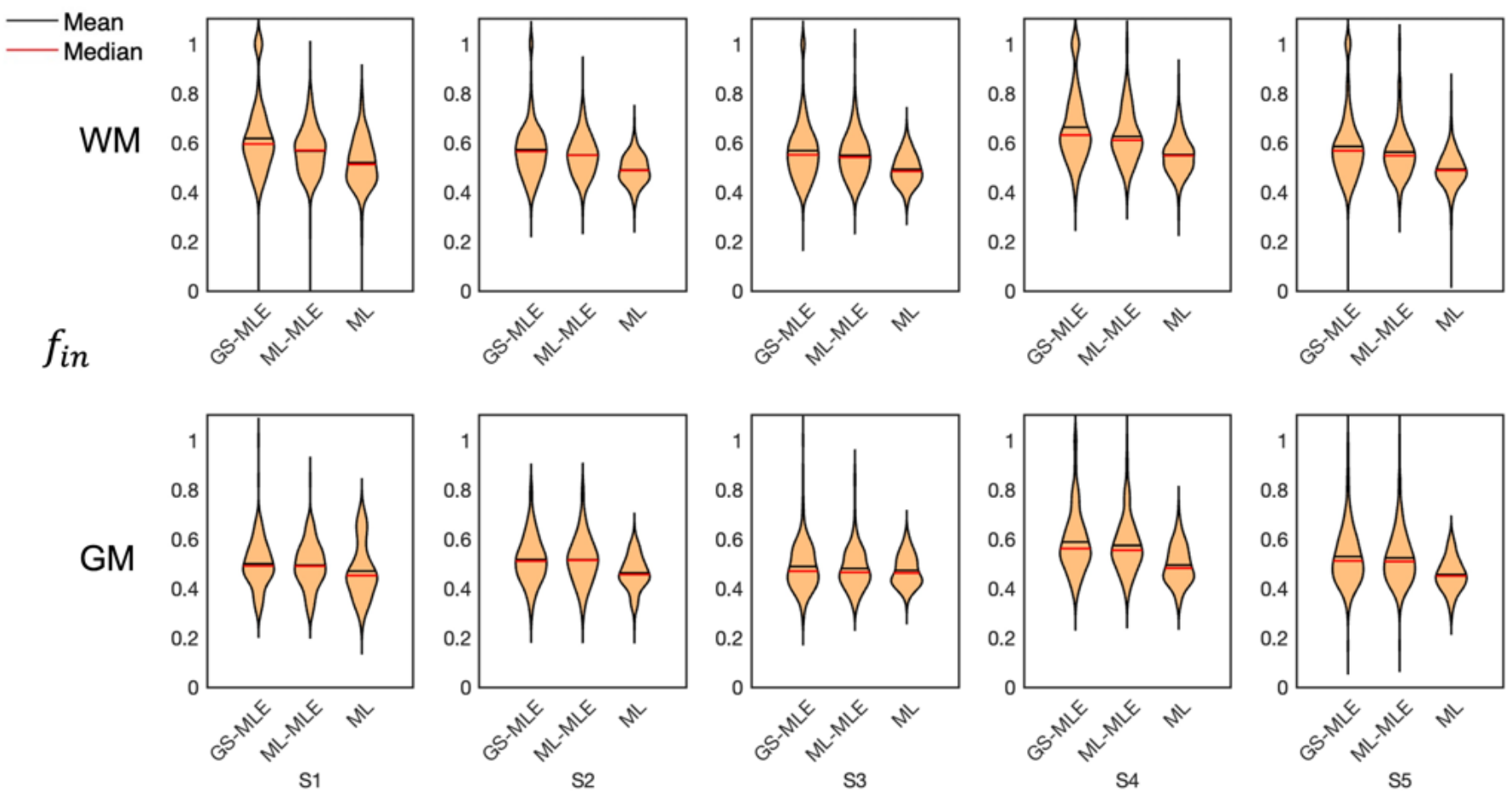

Figure 4. Distributions of $f_{in}$ from all subjects in the WM and GM. Outliers generated from GS-MLE in the WM are observed for all subjects, which are reduced in ML-MLE estimation, giving a closer mean and median in the distribution; ML estimation gives systematically lower $f_{in}$ estimates than the MLE-based method in all subjects.

Table 1. Mean and standard deviation (std) of parameters in WM and GM from all subjects. ML-MLE reduces the std of $f_{in}$ in WM due to the elimination of outlier estimates in GS-MLE.

| $f_{in}$ | WM | | | | | | GM | | | | | |
|---|---|---|---|---|---|---|---|---|---|---|---|---|
| | GS-MLE | | ML-MLE | | ML | | GS-MLE | | ML-MLE | | ML | |
| | mean | std | mean | std | mean | std | mean | std | mean | std | mean | std |
| S1 | 0.617 | 0.149 | 0.566 | 0.102 | 0.519 | 0.102 | 0.499 | 0.100 | 0.494 | 0.093 | 0.470 | 0.112 |
| S2 | 0.571 | 0.114 | 0.549 | 0.083 | 0.489 | 0.060 | 0.516 | 0.098 | 0.516 | 0.097 | 0.462 | 0.070 |
| S3 | 0.567 | 0.123 | 0.548 | 0.094 | 0.491 | 0.063 | 0.488 | 0.098 | 0.480 | 0.087 | 0.473 | 0.064 |
| S4 | 0.662 | 0.143 | 0.625 | 0.107 | 0.551 | 0.080 | 0.588 | 0.126 | 0.574 | 0.111 | 0.494 | 0.079 |
| S5 | 0.585 | 0.141 | 0.562 | 0.106 | 0.492 | 0.076 | 0.528 | 0.118 | 0.523 | 0.111 | 0.455 | 0.068 |

| $f_{iso}$ | WM | | | | | | GM | | | | | |
|---|---|---|---|---|---|---|---|---|---|---|---|---|
| | GS-MLE | | ML-MLE | | ML | | GS-MLE | | ML-MLE | | ML | |
| | mean | std | mean | std | mean | std | mean | std | mean | std | mean | std |
| S1 | 0.167 | 0.157 | 0.163 | 0.136 | 0.131 | 0.093 | 0.106 | 0.115 | 0.104 | 0.112 | 0.068 | 0.051 |
| S2 | 0.089 | 0.113 | 0.081 | 0.099 | 0.113 | 0.058 | 0.047 | 0.076 | 0.048 | 0.074 | 0.068 | 0.036 |
| S3 | 0.101 | 0.123 | 0.098 | 0.111 | 0.104 | 0.060 | 0.060 | 0.088 | 0.058 | 0.080 | 0.060 | 0.038 |
| S4 | 0.145 | 0.144 | 0.132 | 0.134 | 0.150 | 0.083 | 0.091 | 0.112 | 0.087 | 0.108 | 0.095 | 0.062 |
| S5 | 0.157 | 0.150 | 0.154 | 0.140 | 0.129 | 0.077 | 0.103 | 0.121 | 0.104 | 0.116 | 0.084 | 0.061 |

| $ODI$ | WM | | | | | | GM | | | | | |
|---|---|---|---|---|---|---|---|---|---|---|---|---|
| | GS-MLE | | ML-MLE | | ML | | GS-MLE | | ML-MLE | | ML | |
| | mean | std | mean | std | mean | std | mean | std | mean | std | mean | std |
| S1 | 0.030 | 0.044 | 0.021 | 0.050 | 0.013 | 0.047 | 0.108 | 0.075 | 0.103 | 0.079 | 0.082 | 0.080 |
| S2 | 0.035 | 0.032 | 0.026 | 0.032 | 0.015 | 0.019 | 0.146 | 0.091 | 0.143 | 0.094 | 0.116 | 0.100 |
| S3 | 0.027 | 0.024 | 0.020 | 0.023 | 0.012 | 0.013 | 0.073 | 0.054 | 0.067 | 0.056 | 0.056 | 0.047 |
| S4 | 0.053 | 0.068 | 0.044 | 0.071 | 0.039 | 0.097 | 0.116 | 0.072 | 0.108 | 0.078 | 0.077 | 0.071 |
| S5 | 0.057 | 0.092 | 0.045 | 0.086 | 0.035 | 0.101 | 0.138 | 0.116 | 0.133 | 0.117 | 0.104 | 0.118 |

## 5 Discussion & Conclusion

In summary, this study proposes an ML-informed MLE approach to address the challenge of unreliable microstructure parameter estimation under low SNR spinal cord diffusion MRI data. In testing NODDI-derived parameters, the ML-informed method can reduce outlier estimates from conventional MLE and avoid high biases from pure ML estimation. The proposed method also speeds up the computation compared to GS-MLE, making it a promising tool for future applications.

MLE can provide a consistent and efficient approach to parameter estimation problems in diffusion MRI while being sensitive to the choice of starting points. On the positive side, MLE is known to provide unbiased estimates as the sample size increases; different diffusion models and the Rician noise model in DWI data can be considered. However, when the measured signals are very noisy, the variance of estimation can increase, decreasing the precision of estimation. As is shown in our results, by providing starting points from ML estimation, some outlier estimates generated from such low SNR data can be eliminated, hence improved precision of estimates can be achieved.

Our study confirms previous finding that the high precision of direct ML estimation can contain strong biases, especially under low SNR (37). In our cases, when CSF contamination is low, ML can generate relatively accurate $f_{in}$ estimation, but the biases go up quickly when there is CSF contamination; while biases of $f_{in}$ are always negative, the biases of $f_{iso}$ can be both positive and negative. The non-uniform pattern of bias makes ML estimation unpredictable for pathological tissue and can hence hamper its clinical utility.

In finding a starting point for the MLE, the ML estimation, though biased, is likely to find a solution within the basin of attraction of the global minimum with a much shorter computation time than the GS method. For grid search methods, to find such a reliable starting point, the density of the searching grids needs to be increased which will lead to an even longer computation time.

Our method uses a three-layer ANN model trained on simulated data to generate the initial guesses. While the model already gives low mean square errors on training and testing dataset, future work will explore bias and variance trade-offs and improvement of overall estimation performance by further exploring factors like network architecture, training samples distribution, choice of training labels (41), and separate optimisation for each parameter etc.

The proposed ML-MLE method includes direct ML estimation, providing an opportunity to combine them for certain tasks to improve the outcome. While the ML-MLE estimates are less biased, ML estimation gives higher precision in the parameter maps. The performance of some clinical tasks, such as classification, may not depend on parameter-estimation accuracy or precision alone. With a task-driven assessment of parameter estimation (42), we may be able to choose between ML-MLE and ML estimation or combine them together to improve the outcome.

# Supplementary Material

**M1.** The distributions of $f_{iso}$ in the whole cord, cord WM and cord GM from in vivo dataset and the distribution of $f_{iso}$ used in training.

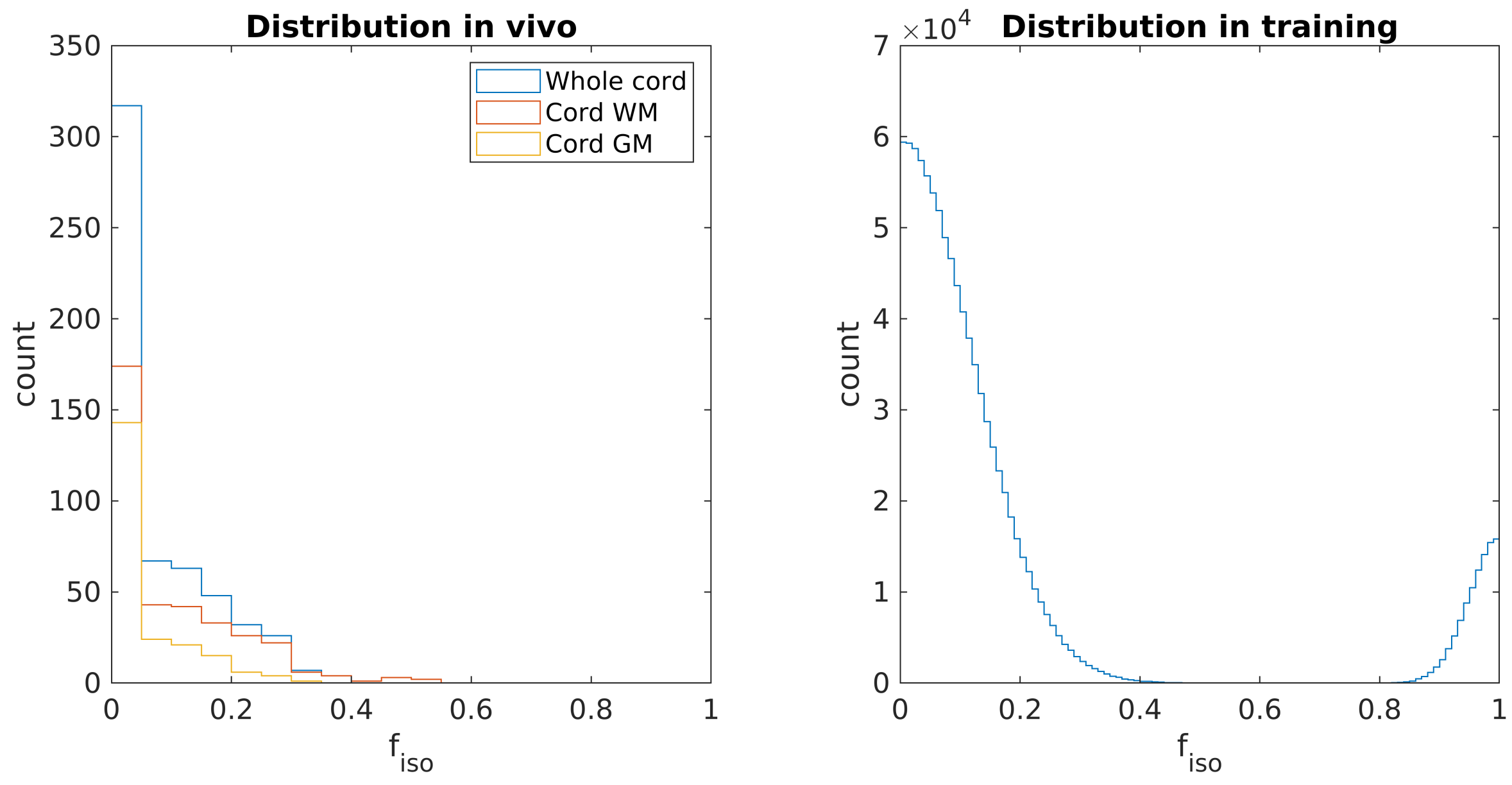

**M2.** The SNR levels for the images of the five in vivo datasets. The signal levels are estimated from the mean signals within the whole cord, cord WM and cord GM from the segmentation on the b=0 images and diffusion-weighted images. The noise levels are estimated from the standard deviations of signals outside the cord body (4 squares in the corners of the background).

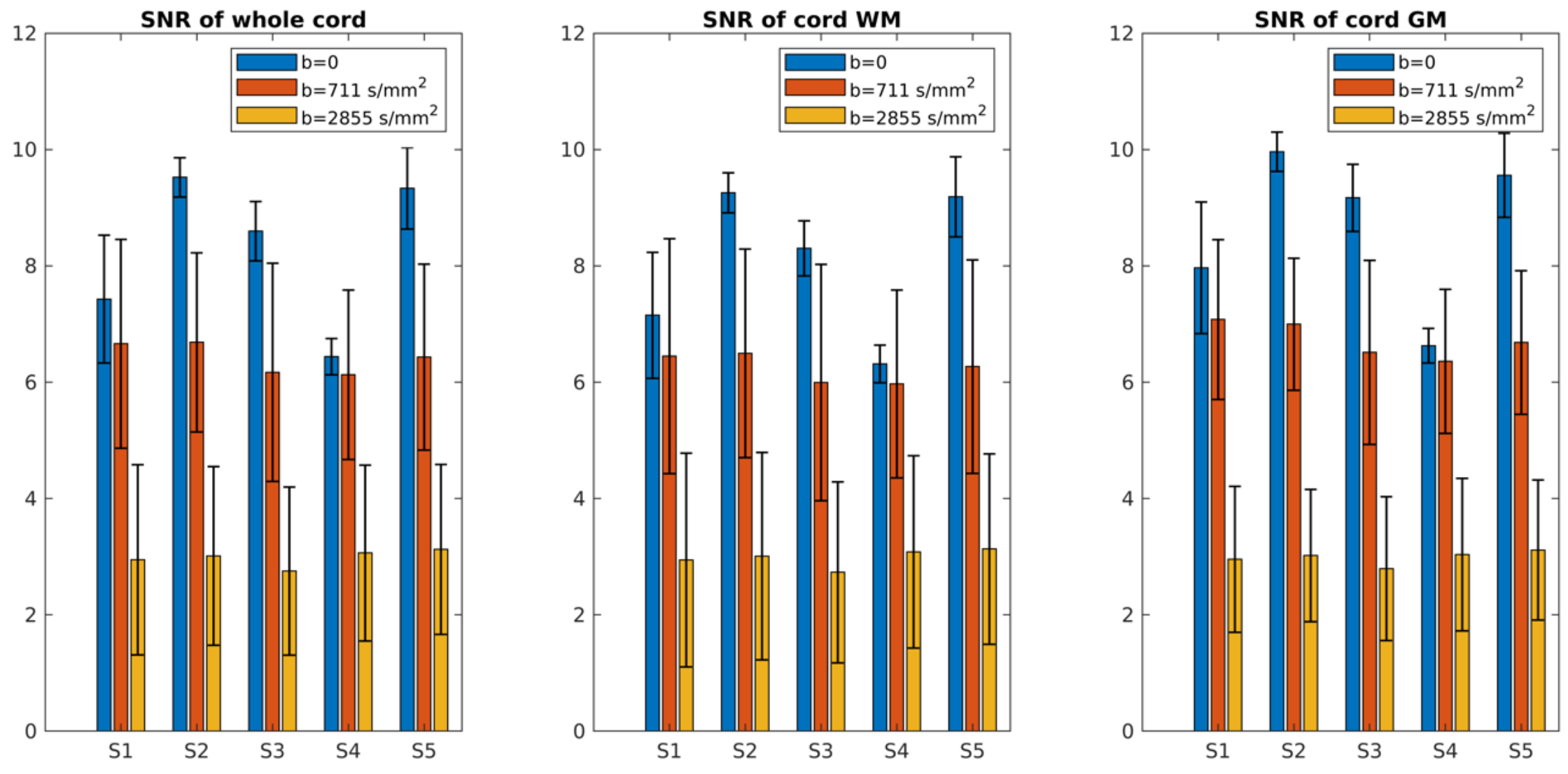

**M3.** 2D distribution of estimated intra-neurite fraction $f_{in}$ and free water fraction $f_{iso}$ from noisy simulations

M3.(A) GS-MLE

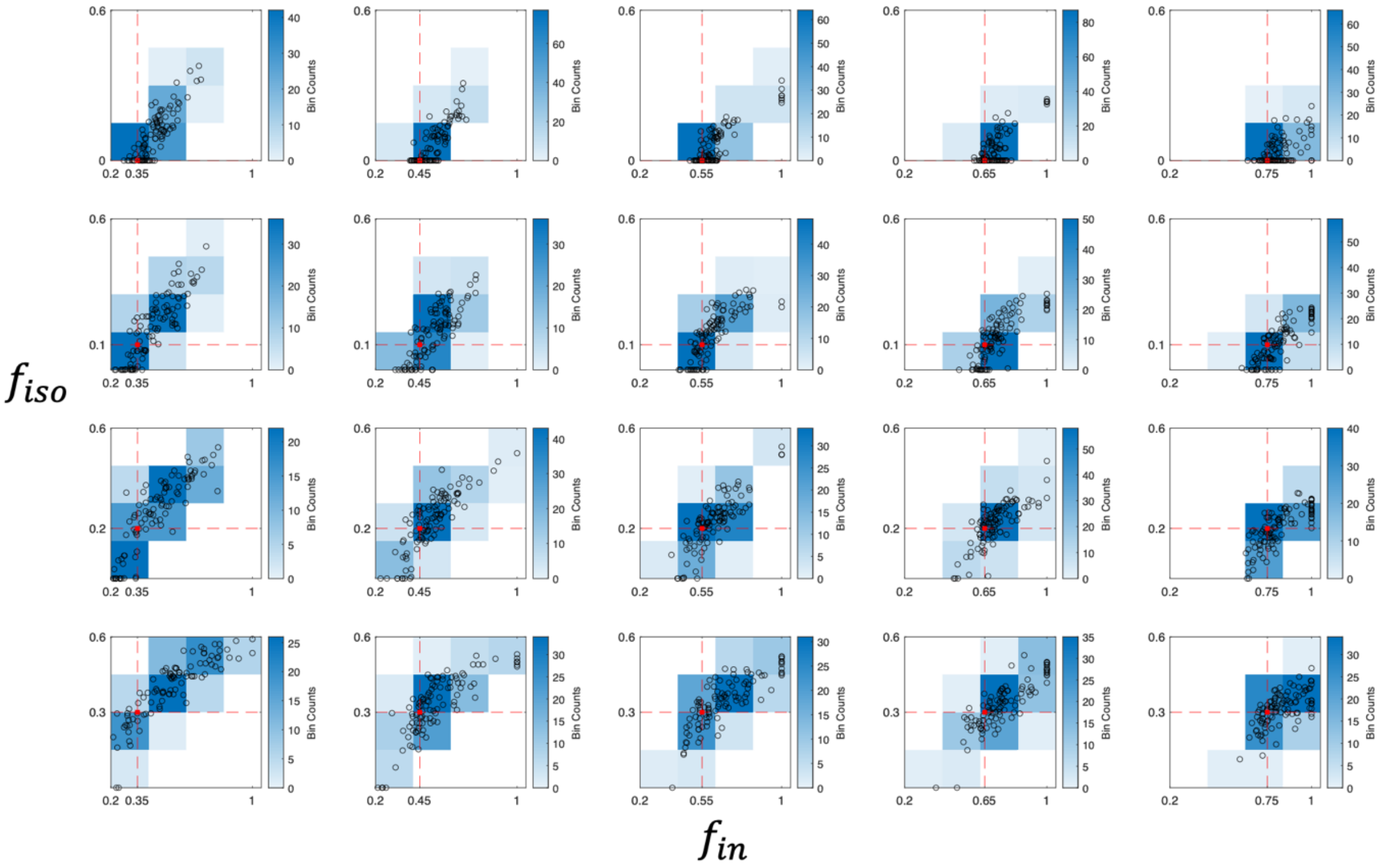

M3.(B) ML-MLE

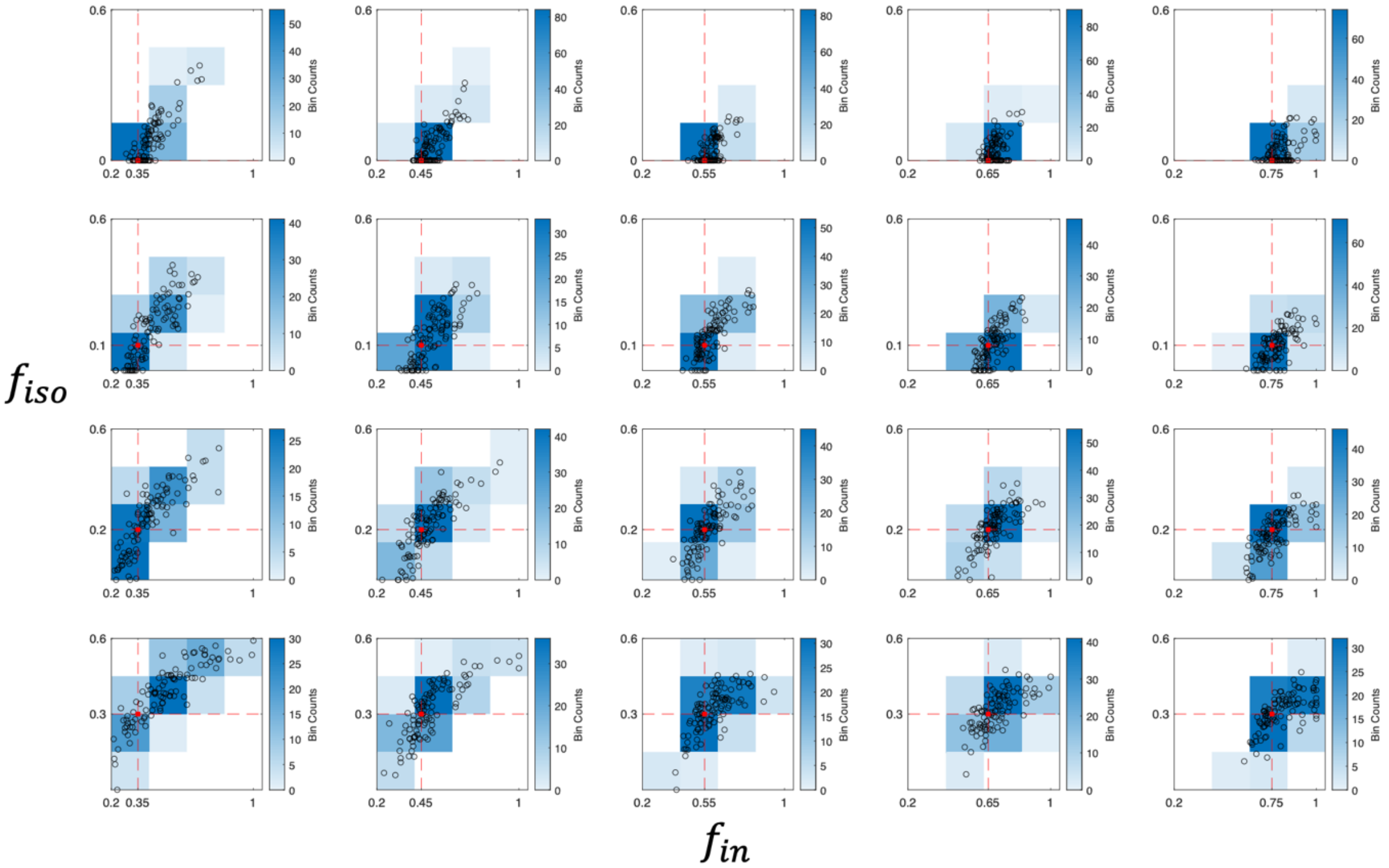

M3.(C) ML

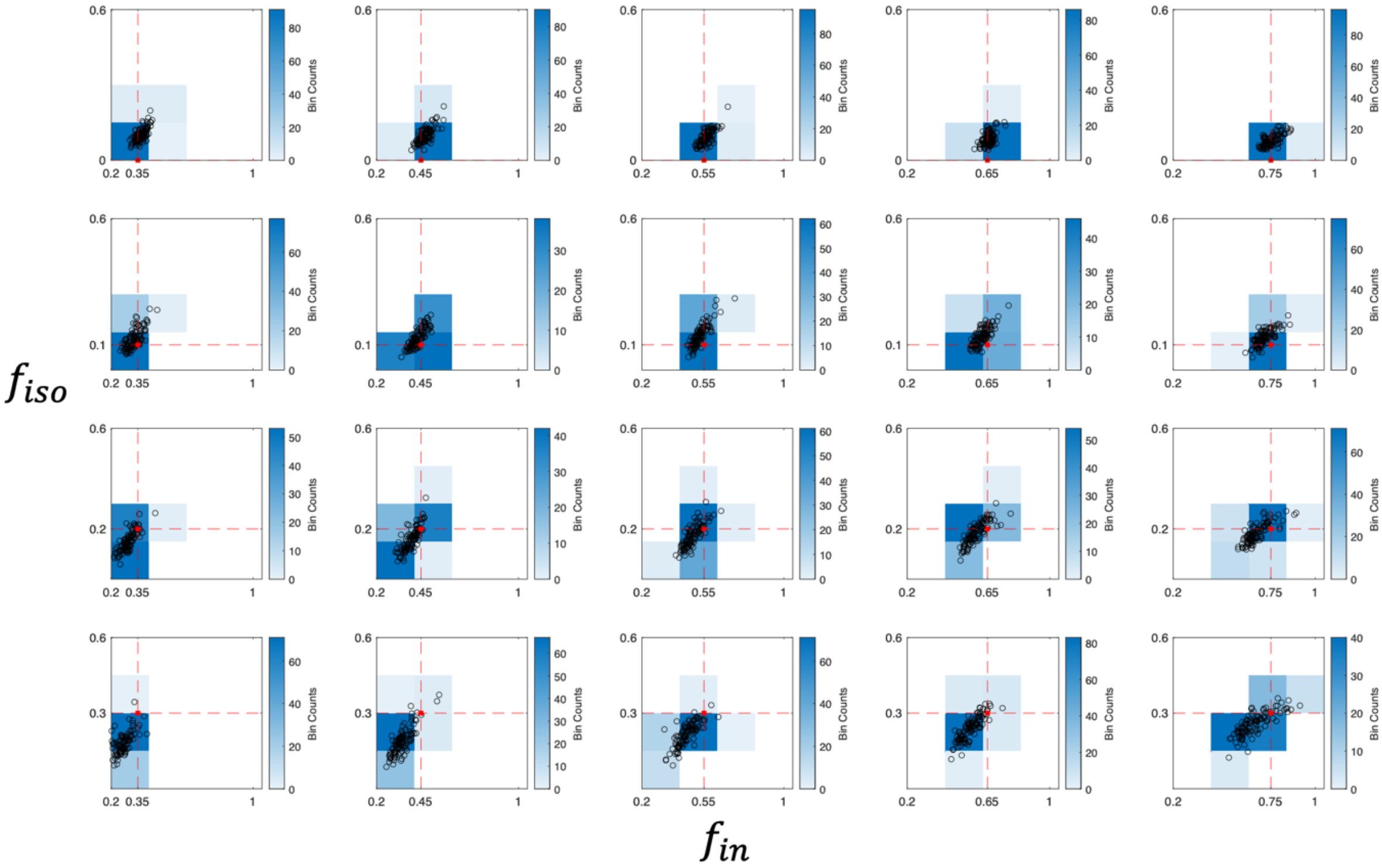

## Acknowledgements

TG is supported by the Medical Research Council (MRC reference: MR/T046473/1). FG is supported by the investigator-initiated PREdICT study at the Vall d'Hebron Institute of Oncology (Barcelona), funded by AstraZeneca. AstraZeneca did not influence data acquisition, analysis, result interpretation and the decision to submit this work in its present form. FG receives the support of a fellowship from the "la Caixa" Foundation (ID 100010434). The fellowship code is "LCF/BQ/PR22/11920010".

The acquisition of the data used in this study was supported by the UCL Grand Challenges scheme, an initiative of the UCL School of Life and Medical Sciences, UCLH/UCL Biomedical Research Centre and the Specialist Biomedical Research Centres at Moorfields/UCL and Great Ormond Street/UCL, as well as by a programme grant from the UK MS Society (grant 892/08). The MRI scanner of the NMR Research Unit, Queen Square Multiple Sclerosis Centre, is supported by the National Institute for Health Research University College London Hospitals Biomedical Research Centre. CGWK receives grant funding from The Multiple Sclerosis Society (grant #77), Wings for Life (#169111), BRC(#BRC704/CAP/CGW), MRC (#MR/S026088/1), and Ataxia UK. CGWK is a shareholder in Queen Square Analytics Ltd.